\documentclass[journal]{IEEEtran}
\usepackage{stfloats}
\usepackage{algorithm}  
\usepackage{algpseudocode}  
\usepackage{amsmath}

\usepackage{graphicx}
\usepackage[]{algpseudocode}
\usepackage{algorithmicx,algorithm}
\usepackage{amsmath}
\usepackage{titlesec}
\usepackage{multirow}  
\usepackage{amsthm,amsmath,amssymb}
\usepackage{graphicx}
\usepackage{float}
\usepackage{subfigure}
\usepackage{mathrsfs}
\usepackage{url}
\usepackage{array}
\newcolumntype{C}[1]{>{\centering}p{#1}}
\usepackage{makecell}
\usepackage{diagbox}
\usepackage{amsmath}
\newtheorem{remark}{Remark}
\usepackage{gensymb}
\usepackage{cite}
\usepackage{color}

\ifCLASSOPTIONcompsoc
 \usepackage[caption=false,font=normalsize,labelfont=sf,textfont=sf]{subfig}
\else
 \usepackage[caption=false,font=footnotesize]{subfig}
\fi
\UseRawInputEncoding
\begin{document}
\newtheorem{theorem}{Theorem}[section] 
\newtheorem{definition}[theorem]{Definition} 
\newtheorem{lemma}{Lemma} 
\newtheorem{corollary}[theorem]{Corollary}
\newtheorem{example}{Example}[section]
\newtheorem{proposition}[theorem]{Proposition}
\vspace{-15mm}
\title{Electromagnetic Property Sensing and Channel Reconstruction Based on Diffusion Schr\"odinger Bridge in ISAC 
}

\author{ 
Yuhua Jiang, Feifei Gao, and Shi Jin

\thanks{
Y. Jiang and F. Gao are with Institute for Artificial Intelligence, Tsinghua University (THUAI), 
State Key Lab of Intelligent Technologies and Systems, Tsinghua University, 
Beijing National Research Center for Information Science and Technology (BNRist), Beijing, P.R. China (email: yh-jiang24@mails.tsinghua.edu.cn, feifeigao@ieee.org).



S. Jin is with the National Mobile Communications Research 
Laboratory, Southeast University, Nanjing 210096, China (e-mail: jinshi@seu.edu.cn).



}
}

\maketitle
\vspace{-15mm}
\begin{abstract}
Integrated sensing and communications (ISAC) has emerged as a transformative paradigm for next-generation wireless systems.
In this paper, we present a novel ISAC scheme that leverages the diffusion Schr\"odinger bridge (DSB) to realize the sensing of electromagnetic (EM) property of a target as well as the reconstruction of the wireless channel. 
The DSB framework connects EM property sensing and channel reconstruction by establishing a bidirectional process: the forward process transforms the distribution of EM property into the channel distribution, while the reverse process reconstructs the EM property from the channel.
To handle the difference in dimensionality between the high-dimensional sensing channel and the lower-dimensional EM property, we generate latent representations using an autoencoder network. 
The autoencoder compresses the sensing channel into a latent space that retains essential features, which incorporates positional embeddings to process spatial context.
The simulation results demonstrate the effectiveness of the proposed DSB framework, which achieves superior reconstruction of the target’s shape, relative permittivity, and conductivity.
Moreover, the proposed method can also realize high-fidelity channel reconstruction given the EM property of the target.
The dual capability of accurately sensing the EM property and reconstructing the channel across various positions within the sensing area underscores the versatility and potential of the proposed approach for broad application in future ISAC systems.
\end{abstract}
\begin{IEEEkeywords}
Electromagnetic (EM) property sensing, channel reconstruction, integrated sensing and communications (ISAC), diffusion Schr\"odinger bridge (DSB), generative artificial intelligence (GAI)
\end{IEEEkeywords}

\IEEEpeerreviewmaketitle

\section{Introduction}
Integrated sensing and communications (ISAC) has recently garnered considerable attention from both academic and industrial experts, particularly due to its promising implications for the sixth-generation (6G) wireless networks \cite{isac9,overview_ourgroup}. Unlike the conventional frequency-division sensing and communications (FDSAC) paradigm, which necessitates distinct frequency bands and infrastructure for each operational function, ISAC enables the simultaneous sharing of time, frequency, power, and hardware resources for both communications and sensing functionalities. It is anticipated that ISAC will surpass FDSAC in terms of spectrum efficiency, energy conservation, and hardware demands \cite{isac1,mypaper3,isac2}. Moreover, ISAC has the potential to be integrated with other innovative technologies, such as reconfigurable intelligent surfaces (RISs), to augment the efficacy of sensing and communications systems \cite{mypaper}. 
Given its myriad advantages, ISAC is expected to play a pivotal role in various emerging applications \cite{v2x,xr,iot}, including digital twins that effectively bridge the physical world with its virtual equivalent in the communications domain \cite{twin}. 
In contrast to image-centric digital twins that prioritize shape and spatial orientation, digital twins designed for communications systems are responsible for the complex reconstruction of communications pathways and the management of channel-specific issues.


Electromagnetic (EM) property sensing represents a groundbreaking advancement in ISAC systems, which leverages the unique property of EM waves to simultaneously sense the environment and enable communications.
This novel approach, as discussed in \cite{mypaper4}, introduces a paradigm shift by using orthogonal frequency division multiplexing (OFDM) signals to acquire the target's EM property and identify the material of the target. 
The integration of multiple base stations (BSs) enhances the performance and accuracy of EM property sensing, as explored in \cite{mypaper5}, where sensing algorithms and pilot are meticulously designed to optimize the sensing process. Additionally, diffusion models have been employed to refine the EM property sensing in ISAC systems, which offers a robust framework to accurately detect and interpret environmental EM characteristics \cite{mypaper6}.

Meanwhile, wireless channel reconstruction is also a critical aspect of modern wireless systems, which enables accurate signal processing and improved system performance. 
Recent advancements, such as deep learning and variational Bayesian methods, have significantly enhanced the accuracy and efficiency of channel reconstruction techniques. For instance, deep plug-and-play priors have been proposed to facilitate multitask channel reconstruction in massive multiple-input multiple-output (MIMO) systems, demonstrating notable improvement in handling complex channel conditions \cite{wan2024deep}. Additionally, variational Bayesian learning has been leveraged to optimize localization and channel reconstruction in RIS-aided systems, offering robust performance in the face of channel uncertainties \cite{li2024variational}. In the realm of MIMO systems with doubly selective channels, diagonally reconstructed channel estimation techniques have been developed to mitigate inter-Doppler interference, ensuring more reliable communications \cite{yin2024diagonally}.
Moreover, near-field MIMO channel reconstruction has shown promise in enhancing the performance of future wireless systems by utilizing limited geometry feedback \cite{eslami2024near}.
Finally, deep learning frameworks have been explored for wireless radiation field reconstruction and channel prediction, pushing the boundaries that can be achieved in wireless systems \cite{lu2024deep}. 

In fact, the fields of EM property sensing and channel reconstruction are intrinsically linked through a shared goal of optimizing the performance and reliability of ISAC systems.
The close relationship between EM property sensing and channel reconstruction suggests that advances in one area could significantly benefit the other, particularly by enhancing the precision and efficiency of data acquisition and processing.
However, no direct attempt has been made to integrate the two fields within a unified framework.


\textcolor{black}{
In this paper, we utilize the diffusion Schr\"odinger bridge (DSB) to realize the sensing of EM property of a target and the reconstruction of the wireless channel in ISAC systems. 
As a powerful tool recently explored in generative artificial intelligence (GAI), DSB offers a framework for transitioning between probability distributions in a controlled manner \cite{sdsb ,dsbm, assgm}. 
The DSB framework connects EM property sensing and channel reconstruction by establishing a bidirectional process: the forward process transforms the distribution of EM property into the channel distribution, while the reverse process reconstructs the EM property from the channel. 
To handle the difference in dimensionality between the high-dimensional sensing channel and the lower-dimensional EM property,
we use an autoencoder network to generate the latent representations of the channel. 
The autoencoder compresses the sensing channel into a latent space that retains essential features, which incorporates positional embeddings to process spatial context.
The latent is then used within the DSB framework to iteratively generate the EM property. 
The simulation results demonstrate the effectiveness of the proposed DSB framework, which achieves superior reconstruction of the target’s shape, relative permittivity, and conductivity.
Moreover, the proposed method can also realize high-fidelity channel reconstruction given the EM property of the target.
The dual capability of accurately sensing the EM property and reconstructing the channel across various positions within the sensing area underscores the versatility and potential of the proposed approach for broad application in future ISAC systems.
}

The rest of this paper is organized as follows. 
Section~\uppercase\expandafter{\romannumeral2} presents the ISAC system model.
Section~\uppercase\expandafter{\romannumeral3} describes the DSB model for EM property sensing and channel reconstruction.
Section~\uppercase\expandafter{\romannumeral4} proposes the approach to generating the latent in DSB.
Section~\uppercase\expandafter{\romannumeral5} provides the numerical simulation results, and 
Section~\uppercase\expandafter{\romannumeral6} draws the conclusion.

Notations: Boldface denotes a vector or a matrix; $j$ corresponds to the imaginary unit; $(\cdot)^H$, $(\cdot)^\top$, and $(\cdot)^*$ represent Hermitian, transpose, and conjugate, respectively; 
$\otimes$ denotes the Kronecker product;
$\mathrm{vec}(\cdot)$ and $\mathrm{unvec}(\cdot)$ denote the vectorization and unvectorization operation; 
$\nabla$ denotes the nabla operator; 
$\mathbf{I}$ denote the identity matrix with compatible dimensions; 
$\left\Vert\mathbf{a}\right\Vert_2$ denotes $\ell_2$-norm of the vector $\mathbf{a}$; 
$\left\Vert\mathbf{A}\right\Vert_F$ denotes Frobenius-norm of the matrix $\mathbf{A}$; 
$\left| \cdot \right|$ denotes the element-wise absolute value of complex vectors or matrices; 
$\Re(\cdot)$ and $\Im(\cdot)$ denote the real and imaginary part of complex vectors or matrices, respectively; 
$\mathscr{P}_N$ denotes the the space of $N$-state path measures on a finite time horizon for any $N \in$ $\mathbb{N}$ in discrete stochastic processes; 
the Kullback-Leibler (KL) divergence between distributions $p$ and $q$ is defined by KL$(p|q)=\int p(x)\log \frac {p(x)}{q(x)}$d$x$;
the distribution of a real-valued Gaussian random vector with mean $\boldsymbol{\mu}$ and covariance matrix $\boldsymbol{\Sigma}$ is denoted as $\mathcal{N} (\boldsymbol{\mu}, \boldsymbol{\Sigma})$;
the distribution of a circularly symmetric complex Gaussian (CSCG) random vector with mean $\boldsymbol{\mu}$ and 
covariance matrix $\boldsymbol{\Sigma}$ is denoted as $\mathcal{C N}(\boldsymbol{\mu}, \boldsymbol{\Sigma})$. 
\begin{figure}[t]
  \centering  \centerline{\includegraphics[width = 6.9cm]{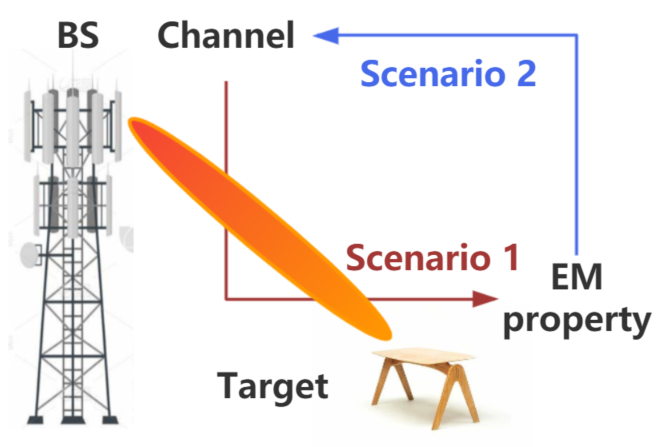}}
  \caption{Multi-antenna mono-static ISAC system for the target's EM property sensing and the channel reconstruction.
  } 
  \label{system_model}
\end{figure}

\section{System Model}
As illustrated in Fig. 1, consider a multi-antenna mono-static ISAC system for the target's EM property sensing and the channel reconstruction. 
The system includes a BS with $N_t$ transmitting antennas and $N_r$ receiving antennas.
We suppose that the BS senses only one target at a time. If there are multiple targets, they can be sensed one by one using a time-division approach.
Since target positioning has been widely studied in ISAC \cite{detection,isac8,zhang2024target}, we assume that the target's location is accurately known by the BS. 
We consider two distinct scenarios in this ISAC system. 
$\textbf{Scenario 1}$: The BS transmits OFDM pilot signals and utilizes the received echo signals to sense the EM property of the target.
$\textbf{Scenario 2}$: The BS is aware of the EM property of the target and reconstruct the OFDM channels without sending any pilot signals.

\subsection{$\textbf{Scenario 1}$}
In the channel estimation stage, the BS adopts fully digital precoding structure where the number of radio frequency (RF) chains $N_{RF}$ is equal to the number of transmitting antennas $N_t$.
The central frequency of the signals is denoted by $f_c$ with the corresponding wavelength $\lambda_c$. 
The number of subcarriers is denoted by $K$ and the frequency spacing between adjacent subcarriers is denoted by $\Delta_f$.
The number of the transmitted symbols in each subcarrier is denoted by $I$.
We assume a quasi-static environment where the channels remain unchanged throughout the sensing period.


Since only the signals scattered by the target carry the information of its EM property, we may send the pilot signals towards the target by beamforming properly at the transmitter.
Let the subscript $k$ represent that the physical quantity is associated with the $k$-th subcarrier.
Denote $\mathbf{H}_{k} \in \mathbb{C}^{N_r \times  N_t}$ as the overall echo channel from the transmitter to the receiver.
Thus, the received signals can be formulated as 
\begin{align}
\mathbf{y}_k = \mathbf{H}_{k} \mathbf{w}_k  +  \mathbf{n}_k ,
\label{y00}
\end{align}
where $\mathbf{w}_k \in \mathbb{C}^{N_t \times  1}$ is the pilot symbol on the $k$-th subcarrier; 
$\mathbf{n}_k \sim \mathcal{C N}\left(\mathbf{0}, \sigma_k ^ 2 \mathbf{I}_{N_r}\right)$ is the CSCG noise at the receiver of the $k$-th subcarrier.
Denote $\mathbf{W}_k =\left[\mathbf{w}_{k,1}, \mathbf{w}_{k,2}, \cdots, \mathbf{w}_{k,I}\right] \in \mathbb{C}^{N_t \times I}$ as the pilot matrix stacked by time,  
and denote $\mathbf{N}_k =\left[\mathbf{n}_{k,1}, \mathbf{n}_{k,2}, \cdots, \mathbf{n}_{k,I}\right] \in \mathbb{C}^{N_r \times I}$. 
Then the overall received pilot signals $\mathbf{Y}_k\in \mathbb{C}^{N_r \times  I}$ can be formulated in a compact form
\begin{align}
\mathbf{Y}_k \triangleq  [\mathbf{y}_{k,1},\mathbf{y}_{k,2},\cdots,\mathbf{y}_{k,I}]
= \mathbf{H}_{k} \mathbf{W}_k + \mathbf{N}_k .    
\label{y1}
\end{align} 

In order to extract the EM property of the target, the BS first needs to estimate $\mathbf{H}_k$ by applying the least square (LS) method as
\begin{align}
\hat{\mathbf{H}}_k = \arg \min_{\mathbf{H}_k} \left\| \mathbf{Y}_k - \mathbf{H}_k \mathbf{W}_k \right\|_F ^2 = \mathbf{Y}_k \mathbf{W}_k^H \left( \mathbf{W}_k \mathbf{W}_k^H \right)^{-1} ,
\end{align}
where $\hat{\mathbf{H}}_k$ is the minimum variance unbiased (MVU) estimated sensing channel. 
We then stack $\hat{\mathbf{H}}_k$ for $ k \in \{1, \cdots , K\} $ into a 3rd order tensor $\hat{\mathcal{H}} \in \mathbb{C} ^ {K \times N_r \times N_t}$ as
\begin{align}
\hat{\mathcal{H}} = \left\{ \hat{\mathbf{H}}_k \right\}_{k=1}^K \in \mathbb{C}^{K \times N_r \times N_t} . 
\label{stack1}
\end{align} 
According to \cite{mypaper4,mypaper5,mypaper6}, the EM property of the target is implicitly encoded in the received echo signals that are transmitted through the sensing channel. 
Thus, we can leverage $\hat{\mathcal{H}}$ as the prior information to reconstruct the EM property of the target. 

\subsection{$\textbf{Scenario 2}$}
Suppose that the EM property of the target exhibits isotropy \cite{li2018deepnis,mesh_free,zhang2022probabilistic}. 
The process of sensing the EM characteristic is essentially about determining the contrast function $\chi_k(\mathbf{r})$, which represents the discrepancy in the complex relative permittivity of the target as compared to the surrounding air. Considering the relative permittivity and conductivity of air to be nearly $1$ and $0$ Siemens per meter (S/m) respectively, the contrast function can be defined as \cite{operator,born2}
\begin{align}
\chi_k(\mathbf{r}) = \epsilon_r(\mathbf{r}) - \frac{j \sigma(\mathbf{r})}{\epsilon_0 \omega_k} - 1 ,
\label{contrast}
\end{align}
where $\epsilon_r(\mathbf{r})$ denotes the real relative permittivity at point $\mathbf{r}$, $\sigma(\mathbf{r})$ denotes the conductivity at point $\mathbf{r}$, $\omega_k = 2 \pi f_k$ denotes the angular frequency of the EM waves, and  $\epsilon_0$ is the vacuum permittivity. 

Throughout this document, it is presumed that the electric fields exhibit a harmonic time variation characterized by $e^{-j \omega_k t}$ \cite{textbook}. 
Let $\lambda_k = c / f_k$ represent the wavelength and $k_k = 2 \pi / \lambda_k$ represent the wave number within the ambient medium.
The total electric field and the incident electric field, which propagate through the medium in the $x$, $y$, and $z$ dimensions, are represented by the complex vectors $\mathbf{E}_k^t ({\mathbf {r}}) \in \mathbb{C}^{3 \times 1}$ and $\mathbf{E}_k^i ({\mathbf {r}}) \in \mathbb{C}^{3 \times 1}$, respectively.
Since the incident electric field is linearly induced by the currents on the transmitting antennas, there is a matrix $\tilde{\mathbf{H}}_{1,k} ({\mathbf {r}}) \in \mathbb{C}^{3 \times N_t}$ that linearly maps 
$\mathbf{w}_k$ to $\mathbf{E}_k^i ({\mathbf {r}})$, i.e.,
\begin{align}
\mathbf{E}_k^i ({\mathbf {r}}) = \tilde{\mathbf{H}}_{1,k} ({\mathbf {r}}) \mathbf{w}_k . 
\label{h1}
\end{align}

Upon exposure to the incident field $\mathbf{E}_k^i ({\mathbf {r}})$, the fields $\mathbf{E}_k^i ({\mathbf {r}})$ and $\mathbf{E}_k^t ({\mathbf {r}})$ are governed by the homogeneous wave equation for the incident field and the inhomogeneous wave equation for the total field, respectively \cite{emscat_polarized}, i.e.,
\begin{align}
\nabla \times \nabla \times \mathbf{E}_k^i ({\mathbf {r}}) - k_k^2
\mathbf{E}_k^i ({\mathbf {r}}) &= \mathbf{0} , \label{inc} \\
\nabla \times \nabla \times \mathbf{E}_k^t ({\mathbf {r}}) - k_k^2 
\mathbf{E}_k^t ({\mathbf {r}}) &= k_k^2 \chi_k(\mathbf{r}) \mathbf{E}_k^t ({\mathbf {r}}) . \label{tot}
\end{align}

Suppose that the BS knows the target is positioned in the region $D$ through prior localization. 
To address the solutions for (\ref{inc}) and (\ref{tot}), the total electric field within $D$ can be formulated by the 3D Lippmann-Schwinger equation \cite{emscat_polarized, lipp, lipp2} 
\begin{equation}
\mathbf{E}_k^t\left(\mathbf{r} \right) = \mathbf{E}_k^i\left(\mathbf{r} \right) + k_k ^2 \iiint_D \overline{\overline{\mathbf{G}}}_k\left(\mathbf{r}, \mathbf{r}^{\prime}\right) \chi_k\left(\mathbf{r}^{\prime}\right) \mathbf{E}_k^t \left(\mathbf{r}^{\prime}\right) \mathrm{d} \mathbf{r}^{\prime},
\label{lipp0}%
\end{equation}
where $\overline{\overline{\mathbf{G}}}_k\left(\mathbf{r}, \mathbf{r}^{\prime}\right) \in \mathbb{C}^{3\times 3}$ is the dyadic electric field Green's function that satisfies
\begin{equation}
\nabla \times \nabla \times \overline{\overline{\mathbf{G}}}_k\left(\mathbf{r}, \mathbf{r}^{\prime}\right) - k_k^2 \overline{\overline{\mathbf{G}}}_k\left(\mathbf{r}, \mathbf{r}^{\prime}\right) = 
\mathbf{I}_3 \delta\left(\mathbf{r}-\mathbf{r}^{\prime}\right) . 
\end{equation}
Meanwhile, $\overline{\overline{\mathbf{G}}}_k\left(\mathbf{r}, \mathbf{r}^{\prime}\right)$ can be formulated as \cite{green}
\begin{align}
&\overline{\overline{\mathbf{G}}}_k\left(\mathbf{r},\mathbf{r}^{\prime}\right)=\left(\mathbf{I}_3+\frac{\nabla \nabla}{k_k^2}\right) g_k\left(\mathbf{r}, \mathbf{r}^{\prime}\right)\nonumber\\
&=\!\!\left[\!\!\left(\frac{3}{k_k^2 R'^2} - \frac{3 j}{k_k R'}-1\!\!\right) \! \hat{\mathbf{r}} \hat{\mathbf{r}}^\top\!\!\!\!-\!\!\left(\frac{1}{k_k^2 R'^2} - \frac{j}{k_k R'}-1\!\!\right)\!\! \mathbf{I}_3\right] \!\!g_k\! \left(\mathbf{r}, \mathbf{r}^{\prime}\right), \label{RR}%
\end{align}
where $R'$ is the distance defined as $R' \triangleq \|\mathbf{r} - \mathbf{r}^{\prime} \|_2$, 
$\hat{\mathbf{r}}\in \mathbb{R}^{3 \times 1}$ is the unit vector from $\mathbf{r}^{\prime}$ to $\mathbf{r}$,
and $g_k\left(\mathbf{r}, \mathbf{r}^{\prime}\right)$ is the scalar Green's function defined as $g_k\left(\mathbf{r}, \mathbf{r}^{\prime}\right) \triangleq \frac{\exp(j k_k R')}{4 \pi R'}$ \cite{green}. 

The echo electric field at the BS's receiver scattered back from the target can then be formulated as \cite{lipp, lipp2}
\begin{equation}
\mathbf{E}_k^s\left(\mathbf{r}_n \right) = k_k ^2 \iiint_D \overline{\overline{\mathbf{G}}}_k\left(\mathbf{r}_n, \mathbf{r}^{\prime}\right) \chi_k \left(\mathbf{r}^{\prime}\right) \mathbf{E}_k^t \left(\mathbf{r}^{\prime}\right) \mathrm{d} \mathbf{r}^{\prime},  
\label{lipp1}%
\end{equation}
where $\mathbf{r}_n$ denotes the position of the $n$-th receiving antenna. 
Suppose the receiver can only measure the scalar electric field component in the direction represented by the unit vector $\mathbf{q}\in \mathbb{R}^{3 \times 1}$. 
The received echo signals can also be formulated as
\begin{align}
\mathbf{y}_k = \tilde{G}_r \left[\mathbf{E}_k^s\left(\mathbf{r}_1 \right) , \cdots, \mathbf{E}_k^s\left(\mathbf{r}_{N_r} \right) \right]^\top \mathbf{q} + \mathbf{n}_k , \label{RR1}%
\end{align}
where $\tilde{G}_r$ denotes the receiving antenna gain. 

\theoremstyle{remark}
\begin{remark} 
In accordance with (\ref{lipp0}), (\ref{lipp1}), and (\ref{RR1}), the echo signals that are transmitted through the sensing channel can also be derived through the EM property of the target. 
As the mapping from $\mathbf{w}_k$ to $\mathbf{y}_k$, $\mathbf{H}_k$ in (\ref{y00}) depends on the EM property of the target and is actually the composite mapping consisting of (\ref{h1}), (\ref{lipp0}), (\ref{lipp1}), and (\ref{RR1}). 
Specifically, (\ref{h1}) maps $\mathbf{w}_k$ to $\mathbf{E}_k^i ({\mathbf {r}})$; 
(\ref{lipp0}) maps $\mathbf{E}_k^i ({\mathbf {r}})$ to $\mathbf{E}_k^t ({\mathbf {r}})$;
(\ref{lipp1}) maps $\mathbf{E}_k^t ({\mathbf {r}})$ to $\mathbf{E}_k^s ({\mathbf {r}})$;
(\ref{RR1}) maps $\mathbf{E}_k^s ({\mathbf {r}})$ to $\mathbf{y}_k$. 
\end{remark}


\section{ DSB for EM Property Sensing and Channel Reconstruction }
\subsection{Point Cloud Representation}

We utilize the point cloud representation to concisely and vividly represent the distribution of the target's EM property.
Define the $m$-th normalized 5D point $\mathbf{x}_m \in \mathbb{R}^{5 \times 1}$ that comprises both the 3D location information and the 2D EM property as
\begin{align}
    \mathbf{x}_m = \left[\frac{x_m - x_c}{x_d}, \frac{y_m - y_c}{y_d}, \frac{ z_m - z_c }{ z_d }, 
    \frac{\epsilon_m - \epsilon_c}{\epsilon_d} , \frac{\sigma_m - \sigma_c}{\sigma_d}  \right]^\top , 
    \label{5D}
\end{align} 
where $x_m$, $y_m$, and $z_m$ denote the coordinates of the $m$-th point in each dimension; 
$x_c$, $y_c$, and $z_c$ denote the coordinates of the center of the target; 
$x_d$, $y_d$, and $z_d$ denote the corresponding standard deviations.
Here, \( \epsilon_m \) and \( \sigma_m \) represent the dielectric constant and conductivity at the \( m \)-th point, respectively; 
\( \epsilon_c \) and \( \sigma_c \) represent their respective central values; 
\( \epsilon_d \) and \( \sigma_d \) denote their respective standard deviations.
Suppose a total of \( M \) points $\mathbf{x}_m$ in (\ref{5D}) constitute the point cloud that represents the target and is defined as \( \mathbf{X}_{\text{data}} \in \mathbb{R}^{M \times 5}\).

The implementation of the point cloud approach presents a more efficient and uncomplicated alternative for discerning the EM property.
Point clouds intrinsically facilitate the distinction between the background medium and the target, thereby eliminating the necessity to analyze the known background medium and significantly reducing the computational burden. Additionally, the representation of data through point clouds permits a clear and prompt visualization of the 3D target, which thereby enhances the intuitive interpretation of the inversion outcomes.

\subsection{DSB Linking EM Property and Sensing Channel}
Let $p_{\text{data}}$ denote the distribution of the EM property and $p_{\text{prior}}$ denote the distribution of the latent extracted from the sensing channel. 
The latent shares the same dimensions as $\mathbf{X}_{\text{data}}$ and is denoted as $\mathbf{X}_{\text{prior}} \in \mathbb{R}^{M \times 5}$. 
To estimate the EM property from the sensing channel or to reconstruct the sensing channel from the EM property, we need to link $p_{\text{data}}$ and $p_{\text{prior}}$ using DSB.
Specifically, DSB establishes the link through a bidirectional process: the forward process gradually transforms $p_{\text{data}}$ into $p_{\text{prior}}$, while the reverse process maps $p_{\text{prior}}$ back to $p_{\text{data}}$. Both processes can be represented via Markov chains.

Denote $\mathbf{X}_{0:N}$ as the set of $ \{ \mathbf{X}_{0}, \cdots ,\mathbf{X}_{N} \}$ that are sequentially generated from $\mathbf{X}_{0} = \mathbf{X}_{\text{data}}$ to $\mathbf{X}_{N} = \mathbf{X}_{\text{prior}}$ over a sequence of \( N - 1 \) intermediate states.
The forward transition \( p_{i+1 \mid i}\left(\mathbf{X}_{i+1} \mid \mathbf{X}_i\right) \) is constructed to progressively guide the distribution from \( p_0 = p_{\text{data}} \) to approximate \( p_N = p_{\text{prior}} \). The joint probability density of \( \mathbf{X}_{0:N} \) is 
\begin{align}
p\left(\mathbf{X}_{0:N}\right) = p_0\left(\mathbf{X}_0\right) \prod_{i=0}^{N-1} p_{i+1 \mid i}\left(\mathbf{X}_{i+1} \mid \mathbf{X}_i\right).
\label{p00}
\end{align}

Similarly, the reverse process can be formulated as a Markovian sequence, with the reverse joint density being
\begin{equation}
q\left(\mathbf{X}_{0:N}\right) = p_N\left(\mathbf{X}_N\right) \prod_{i=0}^{N-1} p_{i \mid i+1}\left(\mathbf{X}_i \mid \mathbf{X}_{i+1}\right),
\label{q00}
\end{equation}
where the conditional probability \( p_{i \mid i+1}\left(\mathbf{X}_i \mid \mathbf{X}_{i+1}\right) \) represents the probability of transitioning from state \( \mathbf{X}_{i+1} \) at time \( i+1 \) to state \( \mathbf{X}_i \) at time \( i \). 
We can decompose the conditional probability $p_{i \mid i+1}\left(\mathbf{X}_i \mid \mathbf{X}_{i+1}\right)$ using Bayesian theorem as
\begin{equation}
p_{i \mid i+1}\left(\mathbf{X}_i \mid \mathbf{X}_{i+1}\right) = \frac{p_{i+1 \mid i}\left(\mathbf{X}_{i+1} \mid \mathbf{X}_i\right) p_i\left(\mathbf{X}_i\right)}{p_{i+1}\left(\mathbf{X}_{i+1}\right)},
\label{Bayesian}
\end{equation}
where \( p_{i+1 \mid i}\left(\mathbf{X}_{i+1} \mid \mathbf{X}_i\right) \) is the conditional probability of the forward process, \( p_i\left(\mathbf{X}_i\right) \) is the marginal distribution of state \( \mathbf{X}_i \), and \( p_{i+1}\left(\mathbf{X}_{i+1}\right) \) is the marginal distribution of state \( \mathbf{X}_{i+1} \).

However, directly computing the conditional probability \( p_{i \mid i+1}\left(\mathbf{X}_i \mid \mathbf{X}_{i+1}\right) \) using (\ref{Bayesian}) is generally quite challenging due to the complexity of the involved distributions and the recursive nature of the computation. In practice, score-based generative models (SGMs) adopt a more tractable approach to handle the forward process, and thus could simplify the forward process by modeling it as the gradual addition of Gaussian noise to the states over time. The forward process can be represented as
\begin{equation}
p_{i+1 \mid i}\left(\mathbf{X}_{i+1} \mid \mathbf{X}_i\right) = \mathcal{N}\left( \mathbf{X}_i + \gamma_{i+1} f_i\left(\mathbf{X}_i\right), 2 \gamma_{i+1} \mathbf{I}\right),
\label{gaussian}
\end{equation}
where \( \gamma_{i+1} \) is the noise level parameter at time step \( i+1 \), and \( f_i\left(\mathbf{X}_i\right) \) is the forward drift term that governs the deterministic part of the state evolution.

For a sufficiently large number of time steps \( N + 1 \), the distribution of the state at the final time step will converge to \( p_{\text{prior}} \), which serves as the starting point for the reverse-time generative process.

The forward process (\ref{p00})
and the reverse process (\ref{q00}) can also be described in a continuous-time framework using stochastic differential equations (SDEs).
Specifically, the forward process can be modeled by an SDE as
\begin{align}
\mathrm{d} \mathbf{X}_t = f_t\left(\mathbf{X}_t\right) \mathrm{d} t + g_t \, \mathrm{d} \mathbf{B}_t,
\label{forw}
\end{align}
where $t \in [0,T]$,  $f_t(\mathbf{X}_t): \mathbb{R}^{M \times 5} \rightarrow \mathbb{R}^{M \times 5}$ is the drift term function, $g_t$ represents the diffusion coefficient, and $\mathbf{B}_t$ denotes the standard Brownian motion.
The reverse process, on the other hand, involves solving the time-reversed version of the SDE in (\ref{forw}), and is given by
\begin{align}
\mathrm{d} \mathbf{X}_t = \left[-f_t\left(\mathbf{X}_t\right) + 2 \nabla_{\mathbf{X}_t} \log p_{T-t}\left(\mathbf{X}_t\right)\right] \mathrm{d} t + g_t \, \mathrm{d} \mathbf{B}_t .
\label{reve2}
\end{align}

\subsection{Iterative Proportional Fitting}
The DSB is an extension of the Schr\"odinger bridge (SB) problem, which incorporates diffusion model to model uncertainty and variability in dynamic systems.
In the context of SB problem, we aim to find an optimal distribution $p^* \in \mathscr{P}_{N+1}$ that minimizes the KL divergence from a reference path measure $p^{\text{ref}} \in \mathscr{P}_{N+1}$. The optimization problem is defined as
\begin{align}
p^* = \operatorname*{argmin}_{p \in \mathscr{P}_{N+1}} \Bigg\{ \mathrm{KL}\left(p \mid p^{\text{ref}}\right) :  p_0 = p_{\text{data}}, p_N = p_{\text{prior}} \Bigg\},
\end{align}
where the marginal distributions $p_0$ and $p_N$ correspond to the distributions at the start and end points of the process, respectively.
Typically, the reference measure path $p^{\text{ref}}$ is generated using the same form of the forward SDE as in (\ref{forw}). 

After determining the optimal solution $p^*$, we can sample $\mathbf{X}_0 \sim p_{\text{data}}$ by first generating $\mathbf{X}_N \sim p_{\text{prior}}$ and then iteratively applying the backward transition $p_{i \mid i+1}\left(\mathbf{X}_i \mid \mathbf{X}_{i+1}\right)$.
Alternatively, we can sample $\mathbf{X}_N \sim p_{\text{prior}}$ by initially drawing $\mathbf{X}_0 \sim p_{\text{data}}$ and subsequently applying the forward transition $p_{i+1 \mid i}\left(\mathbf{X}_{i+1} \mid \mathbf{X}_i\right)$. The SB formulation enables the bidirectional transitions without relying on closed-form expression of $p_{\text{data}}$ and $p_{\text{prior}}$.

Although the SB problem does not have a closed-form solution, it can be tackled using iterative proportional fitting (IPF), which iteratively solves the following optimization problems:
\begin{align}
& p^{2n+1} = \operatorname*{argmin}_{p \in \mathscr{P}_{N+1}} \left\{ \mathrm{KL}\left(p \mid p^{2n}\right):  p_N = p_{\text{prior}} \right\}, \label{joint} \\
& p^{2n+2} = \operatorname*{argmin}_{p \in \mathscr{P}_{N+1}} \left\{ \mathrm{KL}\left(p \mid p^{2n+1}\right): p_0 = p_{\text{data}} \right\} , \label{joint2}
\end{align}
where the superscript $n$ denotes the number of iterations, and the initialization $p^0$ is set as $p^{\text{ref}}$. 
However, implementing IPF in real-world scenarios can be computationally prohibitive, as it requires the calculation and optimization of joint densities.

DSB can be viewed as an approximate method for IPF, which simplifies the optimization of the joint density by decomposing it into a sequence of conditional density optimization tasks. Specifically, the distribution $p$ is split into the forward and backward conditional distributions $p_{i+1 \mid i}$ and $p_{i \mid i+1}$, respectively:
\begin{align}
& p^{2n+1} = 
\operatorname*{argmin}_{p \in \mathscr{P}_{N+1}} \bigg\{ 
\mathrm{KL}\left(p_{i \mid i+1} \mid p_{i \mid i+1}^{2n}\right) 
: p_N = p_{\text{prior}} 
\bigg\}, \label{conditional1} \\
& p^{2n+2} = 
\operatorname*{argmin}_{p \in \mathscr{P}_{N+1}} \bigg\{ 
\mathrm{KL}\left(p_{i+1 \mid i} \mid p_{i+1 \mid i}^{2n+1}\right) 
: p_0 = p_{\text{data}} 
\bigg\}. \label{conditional2}
\end{align}

It can be shown that optimizing conditional distributions $p_{i+1 \mid i}$ and $p_{i \mid i+1}$ in (\ref{conditional1}) and (\ref{conditional2}) leads to the optimization of the joint distribution $p$ in (\ref{joint}) and (\ref{joint2}) \cite{assgm}.
We assume that the conditional distributions $p_{i+1 \mid i}$ and $p_{i \mid i+1}$ are Gaussian distributions, following the assumption commonly used in SGMs and allowing DSB to analytically handle the time reversal process.
Consequently, we employ two separate neural networks to model the forward and backward dynamics.

The forward process, which governs the transition of the state from one time step to the next, is mathematically expressed as
\begin{equation} 
p_{i+1 \mid i}^n(\mathbf{X}_{i+1} \mid \mathbf{X}_i) = \mathcal{N}\left( \mathbf{X}_i + \gamma_{i+1} f_i^n(\mathbf{X}_i), 2 \gamma_{i+1} \mathbf{I}\right) , 
\label{piter}
\end{equation}
where \( p_{i+1 \mid i}^n(\mathbf{X}_{i+1} \mid \mathbf{X}_i) \) denotes the conditional probability distribution of the state \( \mathbf{X}_{i+1} \) given the state \( \mathbf{X}_i \) at the previous time step. In the multivariate Gaussian distribution \( \mathcal{N} \left( \mathbf{X}_i + \gamma_{i+1} f_i^n(\mathbf{X}_i) , 2 \gamma_{i+1} \mathbf{I} \right) \), \( \gamma_{i+1} \) represents the diffusion coefficient that controls the spread of the distribution, and \( f_i^n(\mathbf{X}_i) \) represents the drift term that accounts for the deterministic part of the state transition in the forward process.
For brevity, we define the forward estimation \( F_i^n(\mathbf{X}_i) \) as
\begin{equation}
F_i^n(\mathbf{X}_i) = \mathbf{X}_i + \gamma_{i+1} f_i^n(\mathbf{X}_i).
\label{f_est}
\end{equation}

Conversely, the backward process, which describes the reverse-time evolution of (\ref{piter}), is given by
\begin{equation}
p_{i \mid i+1}^n(\mathbf{X}_i \! \mid \! \mathbf{X}_{i+1}) = 
\mathcal{N}\left(\mathbf{X}_{i+1} + \gamma_{i+1} b_{i+1}^n(\mathbf{X}_{i+1}), 2 \gamma_{i+1} \mathbf{I}\right),
\label{qiter}
\end{equation}
where \( p_{i \mid i+1}^n(\mathbf{X}_i \mid \mathbf{X}_{i+1}) \) represents the conditional probability distribution of the state \( \mathbf{X}_i \) given the state \( \mathbf{X}_{i+1} \) at the subsequent time step.
Similar to the forward process, the backward process is modeled as a multivariate Gaussian distribution \( \mathcal{N}\left( \mathbf{X}_{i+1} + \gamma_{i+1} b_{i+1}^n(\mathbf{X}_{i+1}), 2 \gamma_{i+1} \mathbf{I}\right) \). The term \( b_{i+1}^n(\mathbf{X}_{i+1}) \) in (\ref{qiter}) is the drift term specific to the backward process, 
which plays an analogous role to \( f_i^n(\mathbf{X}_i) \) in the forward process and can be computed according to (\ref{piter}) as
\begin{equation}
b_{i+1}^n\left(\mathbf{X}_{i+1}\right)=-f_i^n\left(\mathbf{X}_{i+1}\right)+2 \nabla_{\mathbf{X}_{i+1}} \log p_{i+1}^n\left(\mathbf{X}_{i+1}\right) .
\label{back1}
\end{equation}
For brevity, we define the backward estimation \( B_i^n(\mathbf{X}_i) \) as
\begin{equation}
B_i^n(\mathbf{X}_i) = \mathbf{X}_i + \gamma_i b_i^n(\mathbf{X}_i) .
\label{back2}
\end{equation}

According to (\ref{piter}) and (\ref{f_est}),
we can compute the probability density function \( p_{i+1}^n(\mathbf{X}_{i+1}) \) at time step \( i + 1 \) as
\begin{align}
p_{i+1}^n(\mathbf{X}_{i+1}) &= ( 4 \pi \gamma_{i+1})^{-\frac{5M}{2}} \int p_i^n(\mathbf{X}_i) \nonumber\\
&\exp\left[-\frac{\|F_i^n(\mathbf{X}_i) - \mathbf{X}_{i+1}\|_F^2}{4\gamma_{i+1}}\right] \mathrm{d}\mathbf{X}_i, \label{eq:pk_propagation}
\end{align}
where the integral normalization factor $( 4 \pi \gamma_i)^{-\frac{5M}{2}}$ comes from $\mathbf{X}_i , \mathbf{X}_{i+1} \in \mathbb{R}^{M\times 5}$.

Taking the logarithm of \( p_{i+1}^n(\mathbf{X}_{i+1}) \) and applying the gradient with respect to \( \mathbf{X}_{i+1} \), we obtain
\begin{align}
\nabla_{\mathbf{X}_{i+1}} \log p_{i+1}^n(\mathbf{X}_{i+1}) &\! = \!\!\! \int \frac{F_i^n(\mathbf{X}_i) - \mathbf{X}_{i+1}}{2\gamma_{i+1}} \nonumber\\
&\times 
p_{i \mid i+1}^n (\mathbf{X}_i|\mathbf{X}_{i+1}) \mathrm{d}\mathbf{X}_i, \label{eq:grad_log_p}
\end{align}
where we employ the Bayesian rule and then use (\ref{piter}) as 
\begin{align}
p_{i \mid i+1}^n (\mathbf{X}_i|\mathbf{X}_{i+1}) & = \frac{p_{i + 1 \mid i}^n (\mathbf{X}_{i+1}|\mathbf{X}_i) p_{i}^n(\mathbf{X}_{i})}{p_{i+1}^n(\mathbf{X}_{i+1})} \nonumber\\
& = \frac{\exp\left[-\frac{\|F_i^n(\mathbf{X}_i) - \mathbf{X}_{i+1}\|_F^2}{4\gamma_{i+1}}\right] p_{i}^n(\mathbf{X}_{i})}{( 4 \pi \gamma_{i+1})^{\frac{5M}{2}}p_{i+1}^n(\mathbf{X}_{i+1})}
.
\end{align}

Substituting (\ref{eq:grad_log_p}) into (\ref{back1}) and utilizing (\ref{f_est}), we obtain 
\begin{align}
b_{i+1}^n(\mathbf{X}_{i+1}) 
&\! = \!\! \!\int \!\! \frac{F_i^n(\mathbf{X}_i) \!-\! F_i^n(\mathbf{X}_{i+1})}{\gamma_{i+1}} p_{i \mid i+1}^n(\mathbf{X}_i|\mathbf{X}_{i+1}) \mathrm{d}\mathbf{X}_i. \label{eq:bias_function}
\end{align}
Now, considering the Bayesian update step and substituting (\ref{eq:bias_function}) into (\ref{back2}),
we can derive the backward estimation \( B_{i+1}^n(\mathbf{X}_{i+1}) \) as
\begin{align}
B_{i+1}^n(\mathbf{X}_{i+1}) &= \mathbb{E} \left[ \mathbf{X}_{i+1} + F_i^n(\mathbf{X}_i) - F_i^n(\mathbf{X}_{i+1}) \mid \mathbf{X}_{i+1} \right], \label{eq:Bayesian_update}
\end{align}
where the expectation is taken over the joint distribution \( p_{i, i+1}^n(\mathbf{X}_i, \mathbf{X}_{i+1}) \).

\begin{figure*}[t]
  \centering  \centerline{\includegraphics[width = 16.9 cm]{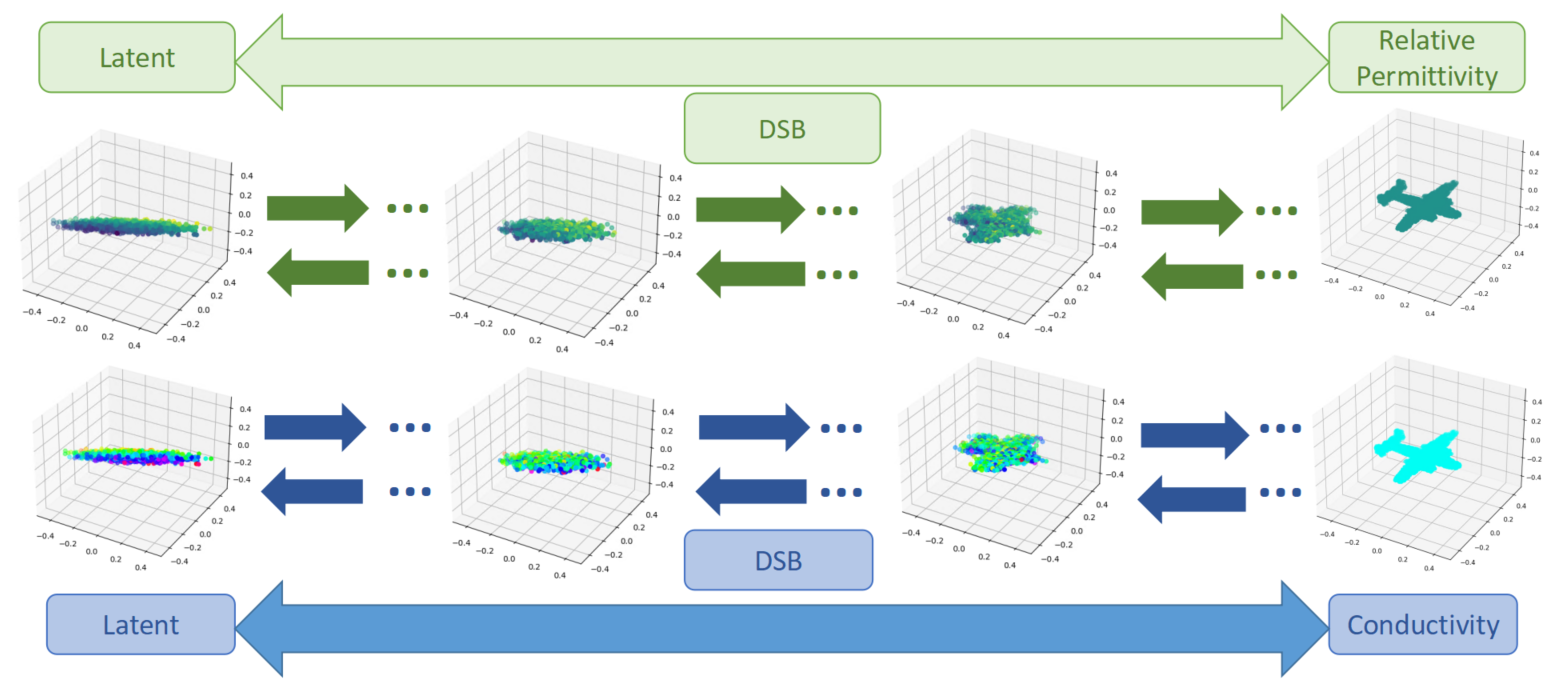}}
  \caption{Schematic diagram of DSB pipeline with forward and backward processes, where the latent refers to the compressed features extracted from the sensing channel.}%
  \label{diffusion_process}%
\end{figure*}

To minimize the difference between $B_{i+1}^n(\mathbf{X}_{i+1})$ and the variable on the right-hand-side of (\ref{eq:Bayesian_update}), we define the loss function \( \mathcal{L}_{B_{i+1}^n} \) as
\begin{align}
\mathcal{L}_{B_{i+1}^n} &= \mathbb{E}_{(\mathbf{X}_i, \mathbf{X}_{i+1}) \sim p_{i, i+1}^n} \Bigg[
\left\|B_{i+1}^n(\mathbf{X}_{i+1}) \right. \nonumber \\
&\quad \left.
- \left(\mathbf{X}_{i+1} + F_i^n(\mathbf{X}_i)\right) 
- F_i^n(\mathbf{X}_{i+1}) \right\|_F^2 \Bigg] .  \label{lossb}
\end{align}
Similarly, the loss function \( \mathcal{L}_{F_i^{n+1}} \) for the mapping function \( F_i^{n+1} \) can be given by
\begin{align}
\mathcal{L}_{F_i^{n+1}} &= \mathbb{E}_{(\mathbf{X}_i, \mathbf{X}_{i+1}) \sim p_{i, i+1}^n} \Bigg[
\left\|F_i^{n+1}(\mathbf{X}_i)  \right. \nonumber \\
&\quad \left.
- \left(\mathbf{X}_i + B_{i+1}^n(\mathbf{X}_{i+1})\right) 
- B_{i+1}^n(\mathbf{X}_i) \right\|_F^2 \Bigg] .  \label{lossf}
\end{align}

The loss functions (\ref{lossb}) and (\ref{lossf}) are derived under the assumption that the distribution \( p_{i, i+1}^n \) accurately models the underlying dynamics of the system and that the error introduced by approximating \( B_{i+1}^n \) and \( F_i^{n+1} \) is minimized in the Frobenius norm sense.
The schematic diagram of DSB pipeline with forward and backward processes 
is shown in Fig.~\ref{diffusion_process}, where the latent refers to the compressed features extracted from the sensing channel. 


In practical applications, the DSB methodology employs two neural networks to approximate the forward estimation (\ref{f_est}) and the backward estimation (\ref{back2}).
Let \( \alpha^n \) and \( \beta^n \) represent the trainable parameters of two neural networks \( F_{\alpha^n}(i, \mathbf{X}_i) \) and \( B_{\beta^n}(i, \mathbf{X}_i) \) in the $n$-th iteration.
Specifically, the neural network \( B_{\beta^n}(i, \mathbf{X}_i) \) is used to approximate the backward estimation \( B_i^n(\mathbf{X}_i) \), 
and the neural network \( F_{\alpha^n}(i, \mathbf{X}_i) \) is used to approximate the forward estimation \( F_i^n(\mathbf{X}_i) \).
The iterative optimization of \( F_{\alpha^n}(i, \mathbf{X}_i) \) and \( B_{\beta^n}(i, \mathbf{X}_i) \) is crucial for the DSB methodology.

\textcolor{black}{
The DSB methodology proceeds by alternately training the backward network \( B_{\beta^n}(i, \mathbf{X}_i) \) and the forward network \( F_{\alpha^n}(i, \mathbf{X}_i) \) across multiple iterations indexed by \( n \).
Specifically, during the \((2n+1)\)-th epoch, we train the backward network \( B_{\beta^n}(i, \mathbf{X}_i) \), while during the \((2n+2)\)-th epoch, we train the forward network \( F_{\alpha^n}(i, \mathbf{X}_i) \).
The alternating optimization of \( B_{\beta^n}(i, \mathbf{X}_i) \) and \( F_{\alpha^n}(i, \mathbf{X}_i) \) is designed to minimize (\ref{conditional1}) and (\ref{conditional2}), which ensures that the forward and backward processes are accurately modeled.
Moreover, both \( B_{\beta^n}(i, \mathbf{X}_i) \) and \( F_{\alpha^n}(i, \mathbf{X}_i) \) are composed of the same concatsquash layers as in \cite{mypaper6}.
}

The convergence of IPF has been rigorously proven in the literature, as demonstrated in \cite{assgm}. The proof underpins the theoretical soundness of the DSB approach, which confirms that the method will, under appropriate conditions, converge to a solution that satisfies the desired characteristics of DSB.

\subsection{DSB Training Scheme}

\begin{algorithm}[t]
\caption{Initial Forward Model Training Scheme}
\begin{algorithmic}[1]
\Require Rounds $R$, timesteps $N$, data distribution $p_{\text{data}}$, prior distribution $p_{\text{prior}}$, and  learning rate $\eta$
\Ensure Initial forward model $F_{\alpha^0}(i, \mathbf{X}_i) = M_\theta(i, \mathbf{X}_i)$
\For{$r \in \{0, \dots, R\}$}
    \While{not converged}
        \State Sample paired data $(\mathbf{X}_0,\mathbf{X}_N)$ with $\mathbf{X}_0 \sim p_{\text{data}}$, $\mathbf{X}_N \sim p_{\text{prior}}$, and $i \in \{0, 1, \dots, N-1\}$
        \State Compute intermediate sample $\mathbf{X}_i = \left(1-\frac{i}{N}\right) \mathbf{X}_0+\frac{i}{N} \mathbf{X}_N $
        \State Compute prediction objective $\mathbf{Y}_i = \mathbf{X}_N-\mathbf{X}_0 $
        \State Compute loss function $\mathcal{L} = \| M_\theta(i, \mathbf{X}_i) - \mathbf{Y}_i \|_F^2$
        \State Update $\theta \gets \theta - \eta \nabla_\theta \mathcal{L}$
    \EndWhile
\EndFor
\State $F_{\alpha^0}(i, \mathbf{X}_i) \gets M_\theta(i, \mathbf{X}_i)$
\end{algorithmic}
\label{algs0}
\end{algorithm}

\subsubsection{Initial Forward Model}
As highlighted in the previous subsection, DSB and SGM are inherently aligned in their training objectives. Specifically, both methodologies aim to model and approximate the target data distribution through a series of learned transformations that progressively refine an initial noise distribution into a more structured and complex distribution that resembles the data.
Given the shared objective of DSB and SGM, it is natural to establish the reference distribution \( p^{\text{ref}} \) in DSB to mirror the noise schedule employed by SGM. By doing so, the training process in the first epoch of DSB becomes theoretically equivalent to the standard training procedure of SGM.

Therefore, rather than initiating the training process of DSB from scratch, we leverage a pre-trained SGM as the starting forward model to train the first backward model.
The pre-trained SGM provides a strong foundation from which DSB can iteratively enhance the quality of generated data through multiple rounds of forward and backward training. Each subsequent epoch in DSB builds upon the results of the previous one, progressively refining the model's ability to capture the complexities of the target distribution. Consequently, DSB develops a new generative model that surpasses the initial SGM in its capacity to accurately model the target data distribution.

The training process for the first forward epoch in DSB, which corresponds to the second epoch in the overall training procedure, is specifically designed to train a neural network to transform the data distribution \( p_{\text{data}} \) into the prior distribution \( p_{\text{prior}} \).
The purpose of the transformation is to ensure that the intermediate states generated during the forward process adhere to the KL divergence constraint, which is imposed by the trajectories learned in the first backward epoch, as outlined in (\ref{conditional1}) and (\ref{conditional2}).


We employ flow matching (FM) models \cite{fm} to train the initial forward model in DSB.
FM models work by ensuring that the flow of the generated data matches the flow of the real data, which effectively captures the underlying dependencies and relationships within the system \cite{fm2}.
By aligning the trajectories of data points, FM models provide a robust framework to model complex systems, which leads to precise and reliable generation \cite{fm3}. 
The entire procedure to train the FM model as the initial forward model within the DSB framework is summarized in Algorithm~\ref{algs0}. 


\begin{figure*}[t]
\centering
\centerline{\includegraphics[width=16.9cm]{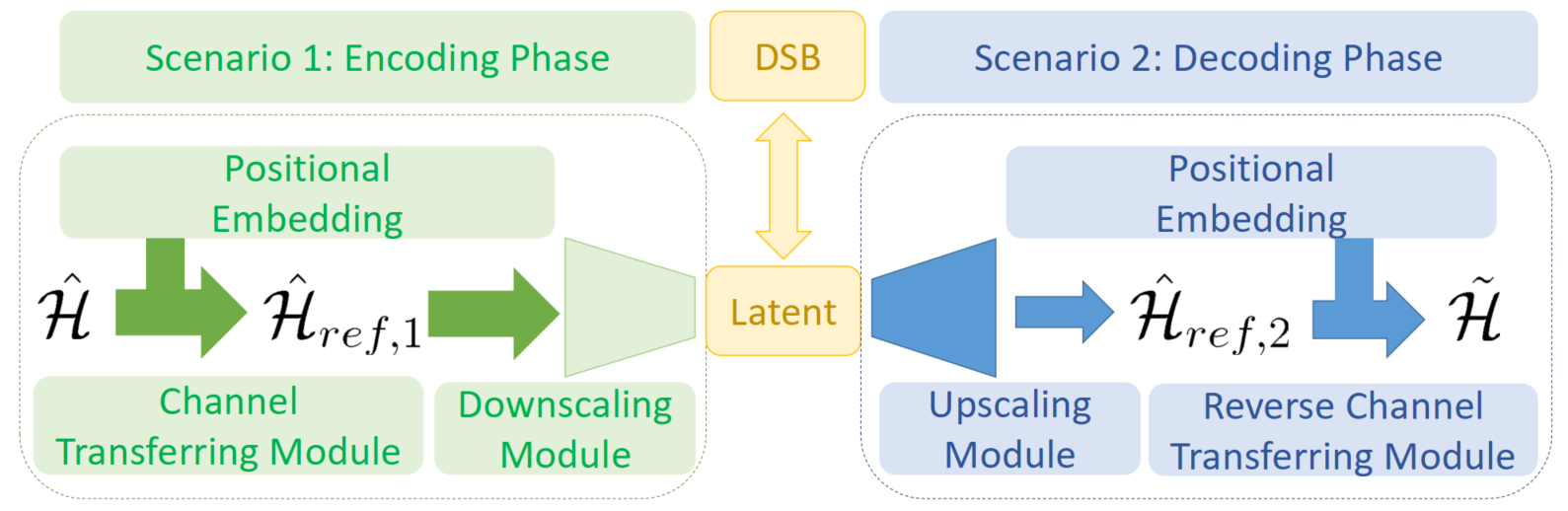}}
  \caption{Schematic diagram of the latent generation using an autoencoder network. The encoding phase is employed in Scenario 1, while the decoding phase is employed in Scenario 2.}
  \label{latent_generation}
\end{figure*}

\begin{algorithm}[t]
\caption{DSB Training Scheme}
\begin{algorithmic}[1]
\Require Epochs $E$, time steps $N$, data distribution $p_{\text{data}}$, and  prior distribution $p_{\text{prior}}$
\Ensure Forward network \( F_{\alpha^{E+1}}(i, \mathbf{X}_{i} ) \) and backward network \( B_{\beta^E}(i, \mathbf{X}_{i} ) \)
\For{$n \in \{0, \dots, E\}$}
    \While{not converged}
        \State Sample $\{\mathbf{X}_i\}_{i=0}^{N}$, where $\mathbf{X}_0 \sim p_{\text{data}}$ and $\mathbf{X}_{i+1} = F_{\alpha^n}(i, \mathbf{X}_i) + \sqrt{2\gamma_{i+1}} \epsilon$, $\epsilon \sim \mathcal{N}(\mathbf{0},\mathbf{I})$  
        \State $ \mathcal{L}_{b_{i+1}^n}' \gets \left\| B_{\beta^n}(i + 1, \mathbf{X}_{i+1}) - \mathbf{X}_i \right\|_F^2$
        \State $\beta^{n+1} \gets \text{Take gradient step}( \nabla_{\beta^n} \mathcal{L}_{b_{i+1}^n}' )$ 
    \EndWhile
    \While{not converged}
        \State Sample $\{\mathbf{X}_i\}_{i=0}^{N}$, where $\mathbf{X}_N \sim p_{\text{prior}}$ and $\mathbf{X}_{i-1} = B_{\beta^n}(i, \mathbf{X}_i) + \sqrt{2\gamma_i} \epsilon$, $\epsilon \sim \mathcal{N}(\mathbf{0},\mathbf{I})$
        \State $ \mathcal{L}_{f_i^{n+1}}' \gets  \left\| F_{\alpha^{n+1}}(i, \mathbf{X}_i) - \mathbf{X}_{i+1} \right\|_F^2$
        \State $\alpha^{n+2} \gets \text{Take gradient step}( \nabla_{\alpha^{n+1}} \mathcal{L}_{f_i^{n+1}}' ) $ 
    \EndWhile
\EndFor
\end{algorithmic}
\label{algs1}
\end{algorithm}
\subsubsection{Loss Function Simplification}
In standard DSB, the loss functions for both the forward and backward models have relatively high computational complexity, as detailed in (\ref{lossb}) and (\ref{lossf}), which renders its physical interpretation challenging. To reduce the complexity, the training loss associated with DSB can be simplified as 
\begin{align}
\mathcal{L}'_{b_{i+1}^n} & = \mathbb{E}_{\left(\mathbf{X}_0, \mathbf{X}_{i+1}\right) \sim p_{0, i+1}^n} \left[ \left\| B_{\beta^n}(i + 1, \mathbf{X}_{i+1}) - \mathbf{X}_i \right\|_F ^2 \right], \label{sb}\\ 
\mathcal{L}'_{f_i^{n+1}} & = \mathbb{E}_{\left(\mathbf{X}_i, \mathbf{X}_N\right) \sim p_{i, N}^n} \left[ \left\| F_{\alpha^{n+1}}(i, \mathbf{X}_i) - \mathbf{X}_{i+1} \right\|_F ^ 2 \right]  \label{sf} . 
\end{align}
The rationale of such simplification will be stated as follows. 
Since generally the drift term $f_i^n\left(\mathbf{X}_i\right)$ changes mildly and $\mathbf{X}_i$ is in close proximity to $\mathbf{X}_{i+1}$ according to \cite{assgm}, 
we can assume $f_i^n\left(\mathbf{X}_i\right) \approx  f_i^n\left(\mathbf{X}_{i+1}\right)$. 
Thus, we have
\begin{align}
& \quad \left[\mathbf{X}_{i+1}+F_i^n\left(\mathbf{X}_i\right)-F_i^n\left(\mathbf{X}_{i+1}\right)\right]-\mathbf{X}_i \nonumber \\
& =  \mathbf{X}_{i+1}+\mathbf{X}_i+\gamma_{i+1} f_i^n\left(\mathbf{X}_i\right) \nonumber \\
& -\left[\mathbf{X}_{i+1}+\gamma_{i+1} f_i^n\left(\mathbf{X}_{i+1}\right)\right] - \mathbf{X}_i \nonumber \\
& =  \gamma_{i+1} f_i^n\left(\mathbf{X}_i\right)-\gamma_{i+1} f_i^n\left(\mathbf{X}_{i+1}\right) \nonumber \\
& \approx  0 . 
\label{app5}
\end{align}
According to (\ref{lossb}) and (\ref{app5}), there is $\mathcal{L}_{B_{i+1}^n}^{\prime} \approx \mathcal{L}_{B_{i+1}^n}$.
Besides, $\mathcal{L}_{F_i^{n+1}}^{\prime} \approx \mathcal{L}_{F_i^{n+1}}$ can also be proved analogously. 
With the simplified loss functions (\ref{sb}) and (\ref{sf}), the overall procedure to train DSB is summarized in Algorithm~\ref{algs1}. 
Once the training of DSB is completed, 
the sampling of DSB can be conducted using (\ref{piter}) and (\ref{qiter}). 

\section{Latent Generation} 
Throughout the DSB process, the dimension of the variables keeps unchanged, which indicates that the dimensions of $\mathbf{X}_{\text{data}}$ and $\mathbf{X}_{\text{prior}}$ should be the same. 
However, in MIMO systems with multiple subcarriers, the dimension of the sensing channel is typically much higher than the dimension of the point cloud that represents the EM property of the target.
Thus, we need to compress the sensing channel into a latent to keep the dimension of the prior equal to the dimension of the 5D point cloud. 

In order to generate the latent with the estimated sensing channel, we adopt an autoencoder network with positional embedding information, as shown in Fig.~\ref{latent_generation}.
The encoding phase is employed in Scenario 1, while the decoding phase is employed in Scenario 2.
The autoencoder is designed to capture the relevant features of the sensing channel \(\mathcal{H}\) while considering the position of the target through positional embedding.

In Scenario 1, the goal is to compress the estimated sensing channel \(\hat{\mathcal{H}}\) into a latent representation that captures the essential features needed for EM property sensing.
First, the estimated sensing channel \(\hat{\mathcal{H}}\) is passed through the channel transferring module, in which 
the positional information of the target is embedded into a high-dimensional 3rd tensor using sinusoidal functions inspired by \cite{attention}.
The positional embedding tensor provides additional context that is crucial for accurate encoding, particularly because the channel features might vary significantly with the target's position. 
The positional embedding tensor is then integrated with the channel and is processed by a neural network.
The resulting combined data 
\(\hat{\mathcal{H}}_{ref,1}\) is then compressed by the downscaling module, which reduces the data size and retains only the most relevant features.
The output of the downscaling module is a latent that shares the same dimensions with the 5D point cloud $\mathbf{X}_{\text{data}}$, i.e., $\mathbb{R}^{M \times 5}$.
The latent is a compressed representation of the target's features, which will be used in DSB to reconstruct the EM property of the target.

In the autoencoder's encoding process, both the channel transferring module and the downscaling module mainly consist of convolutional layers. 
Assume that the dimension of the positional embedding tensor is $\mathbb{R}^{ D_p \times N_r \times  N_t}$. 
The channel transferring module first splits the real and the imaginary parts of the channel and then concatenates them with the positional embedding tensor into the dimension of $\mathbb{R}^{ (2K + D_p) \times N_r \times  N_t}$. 
Next, the downscaling module reduces the spatial dimensions of the data through striding and pooling, which compresses the input into a more compact form. 
The dual role of feature extraction and dimensionality reduction is essential to create an efficient latent representation that retains critical information while minimizing the redundancy.
The last layer of the downscaling module is a fully connected layer that transforms the flattened input into a vector whose length is $5M$ to align with the dimension of the 5D point cloud. 

In Scenario 2, the objective is to reconstruct the channel 
\(\mathcal{H}\) from the latent representation produced by DSB, where we incorporate the same positional information as the encoder to ensure the consistency in the reconstruction.
The latent representation is first passed through the upscaling module, which reverses the compression applied in the encoding phase and expands the latent to \(\hat{\mathcal{H}}_{ref,2}\) with the same size as that of the sensing channel \(\mathcal{H}\). 
Similar to the encoding phase, the same positional embedding tensor is combined with the upscaled data \(\hat{\mathcal{H}}_{ref,2}\), 
which ensures that the positional information is also considered in the decoding phase.
The combined data is then processed by the reverse channel transferring module, which reverts the transformations applied during the encoding phase and aims to reconstruct the sensing channel \(\tilde{\mathcal{H}}\) as accurately as possible.

In the autoencoder's decoding process, the upscaling module reverses the encoding process by taking the flattened latent representation as input and using a fully connected layer followed by a series of transposed convolutional layers.
The reverse channel transferring module concatenates \(\hat{\mathcal{H}}_{ref,2}\) and the positional embedding tensor and then employs a series of convolutional layers to transform the dimension of the data to $\mathbb{R}^{ 2 K \times N_r \times  N_t}$. 
The final output is the reconstructed channel \(\tilde{\mathcal{H}}\) with stacked real and imaginary parts.


Since the magnitude of the sensing channel varies with the location of the target, we need to normalize the loss function.
The training process of the latent generation autoencoder is guided by the normalized mean square error (NMSE) loss function, which measures the difference between the real channel \(\mathcal{H}\) and the reconstructed channel \(\tilde{\mathcal{H}}\) as 
\begin{align}
\mathcal{L}_{\mathrm{NMSE}}&=
\frac{\left\|\mathcal{H}-\tilde{\mathcal{H}}\right\|_F^{2}}{\left\|\mathcal{H}\right\|_F^{2}} .
\label{NMSE}
\end{align} 
By minimizing NMSE of the reconstructed channel \(\tilde{\mathcal{H}}\), the autoencoder learns to produce accurate latent representations that can faithfully reconstruct the sensing channel.

\section{Simulation Results and Analysis}
Suppose all possible targets can be contained within a cubic region $D$ whose size is $1~\mathrm{m}  \times 1~\mathrm{m} \times 1~\mathrm{m}$.
We designate the number of scatter points that form the target as $M = 2048$. 
Assume that the BS is positioned at $(0,0,0)$~m and needs to sense the target in a 30~m radius sector on the horizontal plane, characterized by $ S = \{ ( x , y , 0 ) \mid \operatorname{arctan}\frac{y}{x} \in [-60^\circ, 60^\circ] , \sqrt{x^2+y^2} \leq 30~\mathrm{m} \}$. 
The BS is equipped with a uniform linear array (ULA) with $ N_t = 32 $ transmitting antennas and a ULA with $ N_r = 32 $ receiving antennas. 
The transmitting and the receiving ULAs are both centered at $(0,0,0)$~m and are parallel to $y$ and $z$ directions, respectively, which is analogous to the Mills-Cross configuration \cite{Mills-cross}. 
The transmitting and the receiving antennas are set as dipoles polarized along $z$ and $y$ directions, respectively. 
The central carrier frequency is set as $f_c = 30$~GHz and the corresponding central wavelength is $\lambda_c = 0.01 $~m. 
The inter-antenna spacing for both the transmitting and the receiving ULAs is set as $ \lambda_c/2 = 0.005 $~m. 
We assume there are a total of $K = 16$ subcarriers whose spacing is set as $\Delta_f = 800$~KHz. 

In DSB, we set the number of intermediate time steps as $ N = 100 $. 
The diffusion coefficients $\gamma_i$ linearly increase from $\gamma_0 = 0.001$ to $\gamma_{50} = 0.05 $ and then linearly decrease from $\gamma_{50} = 0.05 $ to $\gamma_{100} = 0.001 $.
In order to train DSB, we select 100000 targets from the ShapeNet dataset \cite{shapenet}, which are split into training, testing, and validation sets by the ratio 80\%, 10\%, and 10\%, respectively.  
All the targets in the dataset are uniformly and randomly located in the sector $S$.
During the training process, we utilize the Adam optimizer and set the batch size as 128.   
In order to compute the forward scattering, we convert (\ref{lipp0}) into a discrete form by the methods of moments (MoM).
Then the unknown total electric field $\mathbf{E}_i^t(\mathbf{r})$ is determined with the stabilized biconjugate gradient fast Fourier transform (BCGS-FFT) technique \cite{BCGS-FFT}.



\begin{figure*}[t]
  \centering
\begin{minipage}[t]{0.33\linewidth}
\subfigure[Target real relative permittivity]{
\includegraphics[width=6cm,height=5.5cm]{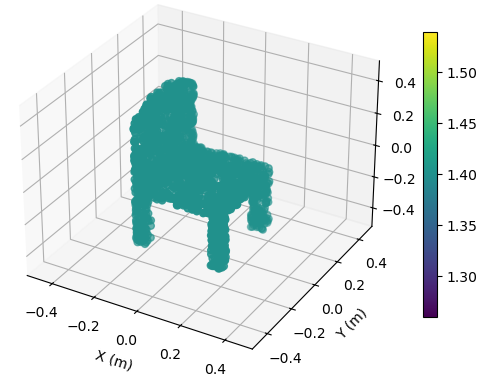}} 
\end{minipage}
\begin{minipage}[t]{0.33\linewidth}
\subfigure[Reconstructed relative permittivity \protect\\ with SNR = 5 dB]{
\includegraphics[width=6cm,height=5.5cm]{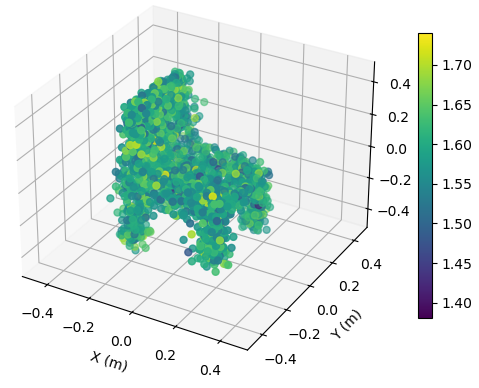}} 
\end{minipage} 
\begin{minipage}[t]{0.32\linewidth}
\subfigure[Reconstructed relative permittivity \protect\\ with SNR = 30 dB]{
\includegraphics[width=6cm,height=5.5cm]{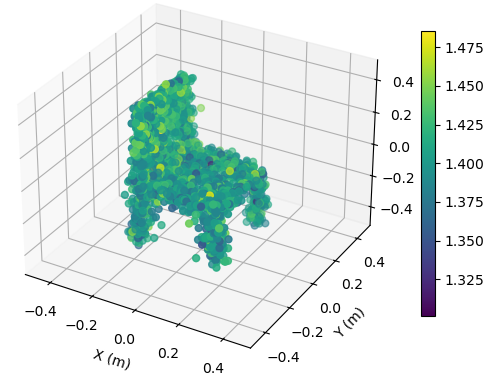}}   
\end{minipage} \\ 
\begin{minipage}[t]{0.33\linewidth} 
\subfigure[Target real conductivity]{
\includegraphics[width=6cm,height=5.5cm]{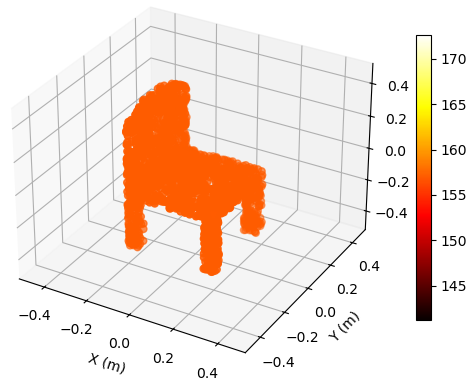}}
\end{minipage}
\begin{minipage}[t]{0.33\linewidth}
\subfigure[Reconstructed conductivity with SNR = 5 dB]{
\includegraphics[width=6cm,height=5.5cm]{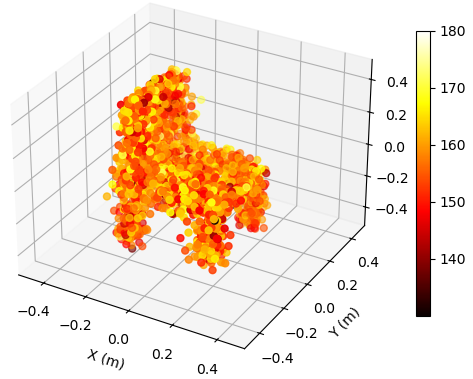}}
\end{minipage}
\begin{minipage}[t]{0.32\linewidth}
\subfigure[Reconstructed conductivity with SNR = 30 dB]{
\includegraphics[width=6cm,height=5.5cm]{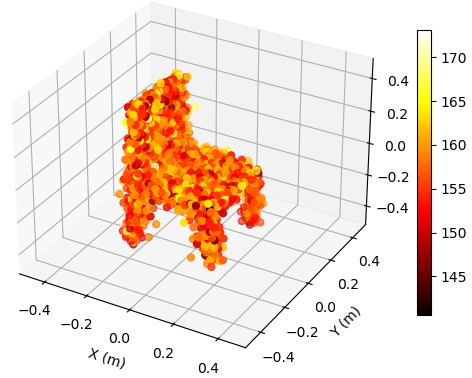}}
\end{minipage}
\caption{ EM property sensing results versus SNR. 
The center of the target is $(15,0,0)$~m. 
The target is shown in the coordinate system relative to its center.
Unit of conductivity is mS/m. } 
\label{image1}
\end{figure*}

Additionally, we propose the mean Chamfer distance (MCD) between the ground truth and the estimated point clouds as a metric to quantitatively assess the performance of EM property sensing, which is defined as
\begin{align}
\mathrm{MCD} \! & = \! 10 \log_{10} \left[\ \frac{1}{| \mathcal{T}|} \sum_{\mathbf{X}_{\text{data}} \in \mathcal{T}} \left(  \frac{1}{M} \sum_{\mathbf{x} \in \mathbf{X}_{\text{data}}} \min _{\mathbf{y} \in \hat{\mathbf{X}}_{data}} \|\mathbf{x} - \mathbf{y}\|_2^2 \! \right. \right. \nonumber\\
& \left.\left. + \frac{1}{M} \sum_{\mathbf{y} \in \hat{\mathbf{X}}_{data}} \min _{\mathbf{x} \in \mathbf{X}_{\text{data}}}\|\mathbf{x}-\mathbf{y}\|_2^2 \right) \right],
\label{CD}%
\end{align} 
where $\mathcal{T}$ denotes the test dataset, $ | \mathcal{T} | $ denotes the number of samples in the test dataset, and $\hat{\mathbf{X}}_{data}$ denotes the estimated value of $\mathbf{X}_{\text{data}}$.

\begin{figure}[t]
  \centering
\centerline{\includegraphics[width=8.4cm,height=6.5cm]{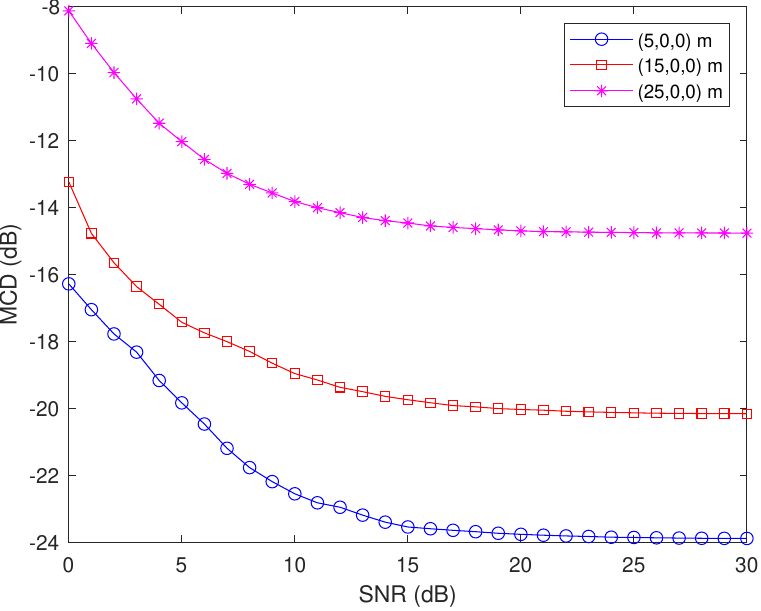}}
  \caption{MCD of 5D point clouds versus SNR. The center of the target is $(15,0,0)$~m.}
  \label{mcd_snr}
\end{figure}

\subsection{  Performance of EM Property Sensing }
In Scenario 1, a total of $ I = 256 $ pilot symbols are transmitted in each subcarrier to estimate the sensing channel.
The signal-to-noise ratio (SNR) at the receiver decides the accuracy of the estimated sensing channel.
The estimated channel is then compressed to generate the latent, which is employed to estimate the EM property of the target.

\subsubsection{ Performance versus SNR at the Receiver}
To illustrate the EM property sensing results, we present the reconstructed 5D point cloud of the target in Fig.~\ref{image1}. 
The center of the target is $(15,0,0)$~m, and the target is shown in the coordinate system relative to its center. 
It is seen from Fig.~\ref{image1} that, the reconstructed point clouds can reflect the general shape of the target. 
The values of EM property reconstructed with SNR = 30 dB is much more accurate compared to those reconstructed with SNR = 5 dB. 
Moreover, a higher SNR value leads to a more precise reconstruction of the target's shape.

We explore the MCD of the reconstructed 5D point clouds versus SNR in Fig.~\ref{mcd_snr}. 
We set the center of the target as $(5,0,0)$~m, $(15,0,0)$~m, or $(25,0,0)$~m, respectively. 
It is seen from Fig.~\ref{mcd_snr} that, the MCD decreases with the increase of SNR to an error floor. 
The MCD is larger when the target is farther from the BS. 
The phenomenon can be attributed to the fact that when the target is closer to the BS, the sensing channel benefits from a higher number of effective degrees of freedom (EDoF) \cite{myEDOF}. As a result, more diverse spatial features of the estimated sensing channel can be extracted, leading to a more accurate reconstruction of the point clouds. Consequently, the error floor of MCD is significantly lower when the target is near the BS, whereas the error increases as the distance between the target and the BS grows, reflecting the reduced EDoF and less diverse spatial feature extraction capability at greater distances.

\subsubsection{ Performance versus Location of the Target }
\begin{figure*}[t]
\captionsetup[subfigure]{singlelinecheck=false} 
\begin{minipage}[t]{0.33\linewidth}
\captionsetup[subfigure]{singlelinecheck=false}
\subfigure[Target relative permittivity]{
\includegraphics[width=6cm,height=5.5cm]{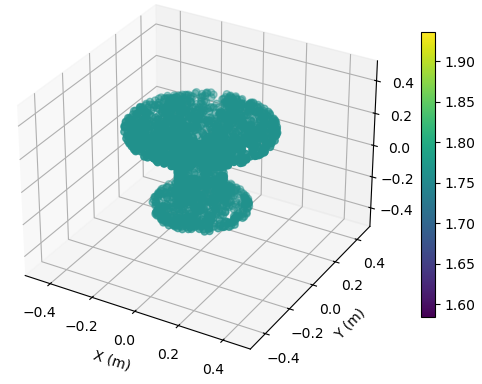}} 
\end{minipage}
\begin{minipage}[t]{0.33\linewidth}
\subfigure[Reconstructed relative permittivity with target \\ at (25,0,0) m ]{
\includegraphics[width=6cm,height=5.5cm]{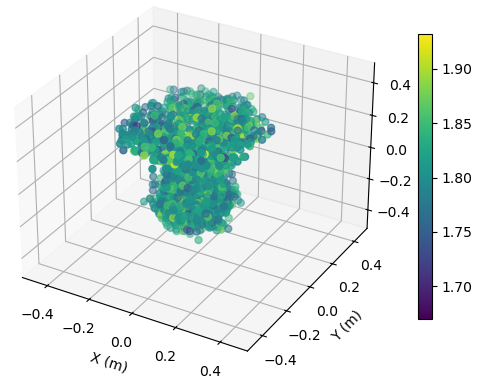}}  
\end{minipage}  
\begin{minipage}[t]{0.32\linewidth}
\subfigure[Reconstructed relative permittivity with target \\at (5,0,0) m ]{
\includegraphics[width=6cm,height=5.5cm]{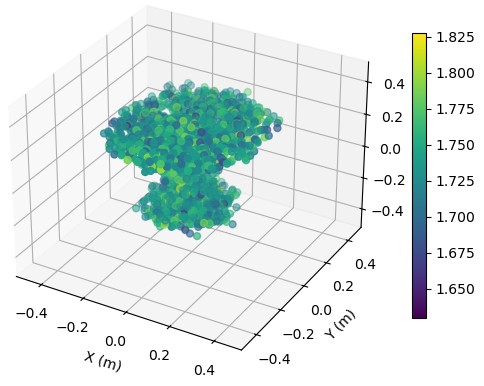}}   
\end{minipage} \\              
\begin{minipage}[t]{0.33\linewidth}
\captionsetup{singlelinecheck=false}
\subfigure[Target conductivity]{
\includegraphics[width=6cm,height=5.5cm]{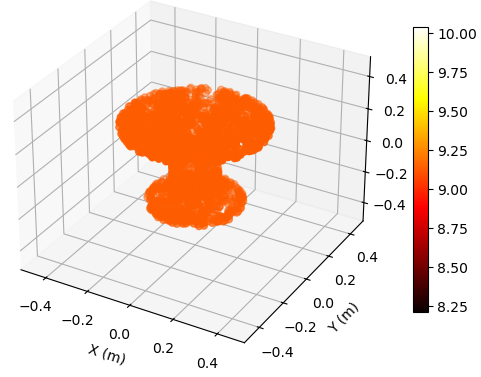}}
\end{minipage}
\begin{minipage}[t]{0.33\linewidth}
\subfigure[Reconstructed conductivity with target \\at (25,0,0) m ]
{\includegraphics[width=6cm,height=5.5cm]{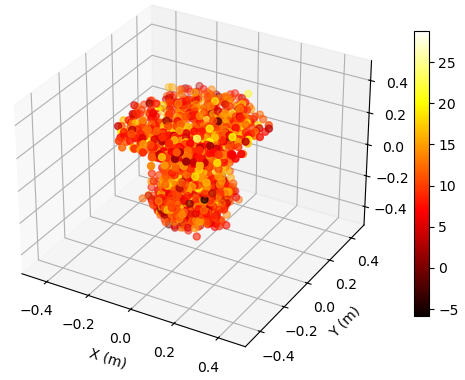}}
\end{minipage}
\begin{minipage}[t]{0.32\linewidth}
\subfigure[Reconstructed conductivity with target \\at (5,0,0) m ]{
\includegraphics[width=6cm,height=5.5cm]{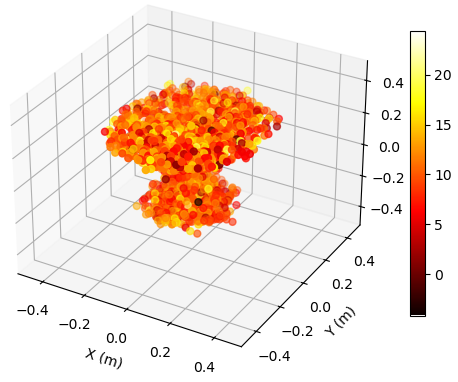}}
\end{minipage}
\caption{EM property sensing results versus location of the target with SNR = 15 dB. 
The target is shown in the coordinate system relative to its center.
Unit of conductivity is mS/m. 
}
\label{image_location}
\end{figure*}

\begin{figure}[t]
  \centering
\centerline{\includegraphics[width=8.4cm,height=6.5cm]{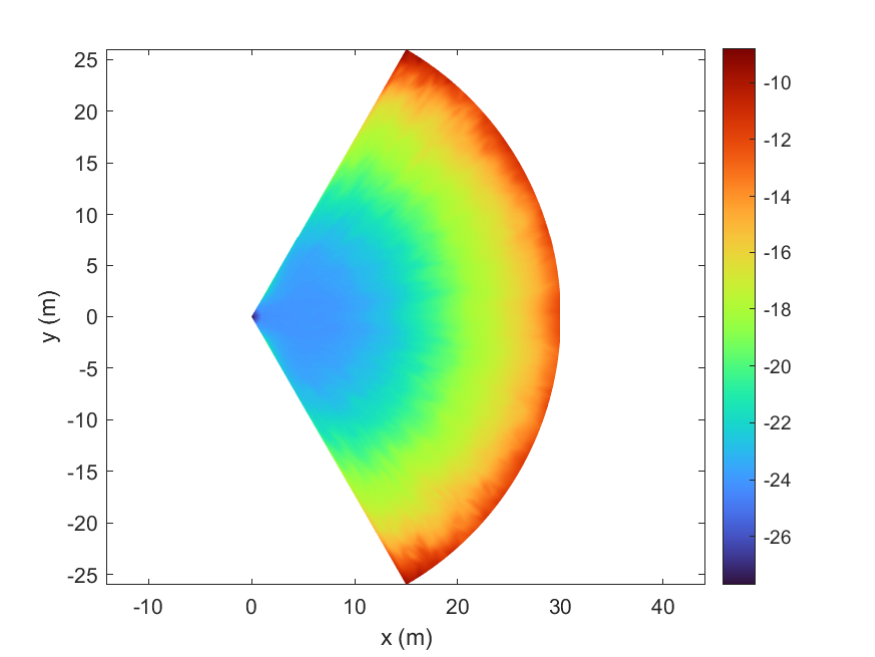}}
  \caption{MCD of 5D point clouds versus location of the target, with SNR = 30 dB.}
  \label{mcd_location}
\end{figure}

To illustrate the EM property sensing results, we present the reconstructed point cloud of the target with SNR = 15 dB in Fig.~\ref{image_location}.
The target is shown in the coordinate system relative to its center. 
It is seen from Fig.~\ref{image_location} that, the reconstructed 5D point clouds can reflect the general shape of the target. 
The values of EM property reconstructed with target at $(5,0,0)$~m is more accurate compared to those reconstructed with target at $(25,0,0)$~m. 
Moreover, a closer distance results in a more accurate reconstructed shape of the target. 

We investigate the MCD of the reconstructed 5D point clouds in relation to the target's location with an SNR of 30 dB, as illustrated in Fig.~\ref{mcd_location}. The figure reveals that the MCD tends to be lower when the target is positioned closer to the BS. The observation can be explained by the fact that a closer target results in a sensing channel with a greater number of EDoF \cite{myEDOF}, allowing for the extraction of more diverse spatial features from the estimated sensing channel.
Additionally, the MCD shows minimal variation with changes in the angle, indicating that the proposed method is capable of effectively sensing the EM property of the target from any direction within the sector $S$.

\subsection{  Performance of Channel Reconstruction }
In Scenario 2, the sensing channel is reconstructed given the EM property and the location of the target. 
We assume the EM property may not be accurate due to measurement errors.

\subsubsection{ Performance versus SNR of EM Property }
\begin{figure}[t]
  \centering
\centerline{\includegraphics[width=8.4cm,height=6.5cm]{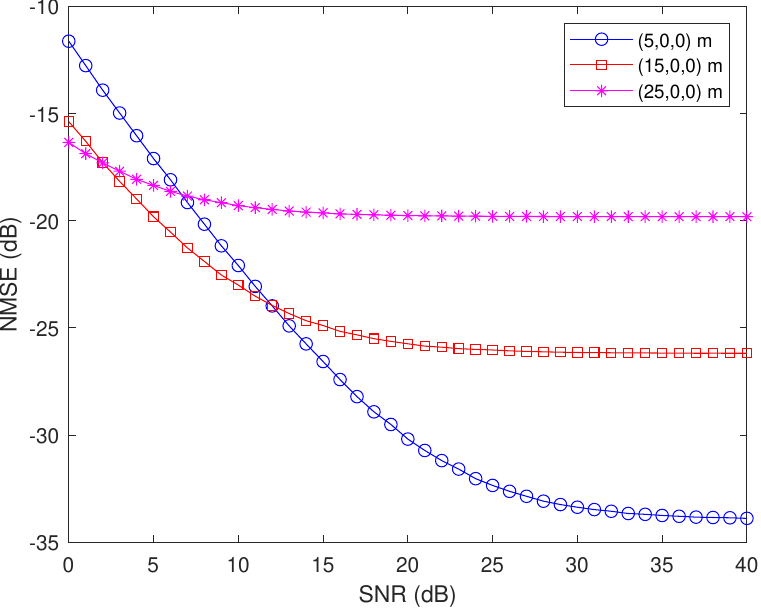}}
  \caption{NMSE of channel reconstruction versus SNR of EM property.}
  \label{nmse_snr}
\end{figure}

We add Gaussian noise to the 5D point cloud that represents the real EM property of the target, and explore the NMSE of channel reconstruction versus SNR of the EM property.
It is seen from Fig.~\ref{nmse_snr} that, as the SNR increases, the NMSE decreases for all target locations.
The error floor reaches approximately -34 dB for the target at $(5,0,0)$~m, around
-27 dB for the target at $(15,0,0)$~m, and about -20 dB for the target at $(25,0,0)$~m. 
However, in the low SNR regions, there is an abnormal phenomenon where the NMSE for the target at $(25,0,0)$~m is smaller than NMSE for targets at $(15,0,0)$~m and $(5,0,0)$~m. This behavior suggests that when the target is farther from the BS, the reconstructed channel becomes more dependent on the target's location and less dependent on its EM property. As a result, the reconstructed channel is less sensitive to the noise affecting the EM property, which leads to higher NMSE values at lower SNR levels compared to the closer targets.

\subsubsection{ Performance versus Location of the Target }
\begin{figure}[t]
  \centering
\centerline{\includegraphics[width=10.4cm,height=7.5cm]{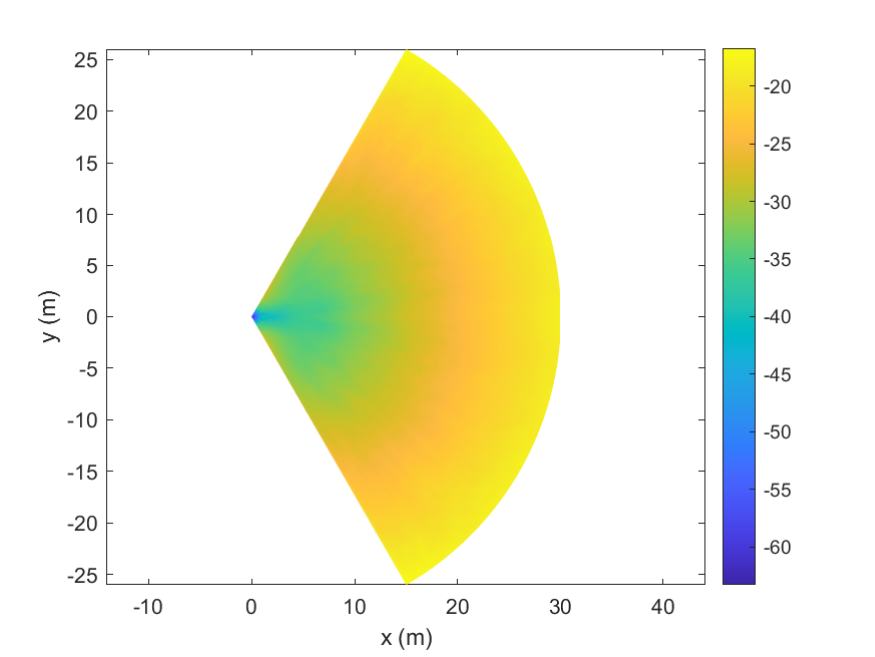}}
  \caption{NMSE of channel reconstruction versus location of the target with accurately known EM property.}
  \label{nmse_location}
\end{figure}

We explore the NMSE of channel reconstruction versus location of the target with accurately known EM property in Fig.~\ref{nmse_location}. 
The NMSE values range from approximately 
-60 dB near the BS to about 
-20 dB at the farthest points in the sector. 
The trend indicates that the NMSE decreases as the distance from the BS increases, yet the variation in angle is not significant.
This trend suggests that the channel reconstruction generally becomes more reliable when the target is closer to the BS.

\section{Conclusion}

This paper introduces a cutting-edge ISAC scheme that utilizes DSB to Bayesian EM property sensing and channel reconstruction within a specific area. The DSB framework facilitates a bidirectional transformation, converting the sensed EM property distribution into a channel distribution and vice versa, while an autoencoder network addresses the dimensionality discrepancy by creating latent representations that maintain crucial spatial features.
The latent representations are then used in DSB to progressively generate the EM property of the target.
Simulation results highlight the superiority of the DSB framework in reconstructing the target's shape, relative permittivity, and conductivity.
Besides, the proposed method is capable of achieving precise channel reconstruction based on the EM property of the target.
The method's ability to accurately detect the EM property and reconstruct channels at different locations within the sensing region highlights its adaptability and promise for widespread use in the ISAC systems.

 \small 
 \bibliographystyle{ieeetr}
 \bibliography{IEEEabrv,mainbib}

\ifCLASSOPTIONcaptionsoff
  \newpage
\fi  



%

\end{document}